\documentclass[aps,prb,groupedaddress,twocolumn]{revtex4-1}
\usepackage{graphicx}
\usepackage{float}

\begin{document}

\title{Measuring the superconducting coherence length in thin films using a two-coil experiment}

\author{John Draskovic}
\email[]{draskovic.1@osu.edu}
%\homepage[]{Your web page}
%\thanks{}
%\altaffiliation{}
\affiliation{Department of Physics, The Ohio State University, Columbus, Ohio, 43210, USA}

\author{Thomas R. Lemberger}
\affiliation{Department of Physics, The Ohio State University, Columbus, Ohio, 43210, USA}
\author{Jaseung Ku}
\affiliation{Department of Physics, University of Illinois at Urbana-Champaign, Urbana, Illinois 61801, USA}
\author{Alexey Bezryadin}
\affiliation{Department of Physics, University of Illinois at Urbana-Champaign, Urbana, Illinois 61801, USA}
\author{Brian Peters}
\affiliation{Department of Physics, The Ohio State University, Columbus, Ohio, 43210, USA}
\author{Fengyuan Yang}
\affiliation{Department of Physics, The Ohio State University, Columbus, Ohio, 43210, USA}
\author{Song Wang}
\affiliation{Department of Physics, Stanford University, Stanford, CA, 94305, USA}

\date{\today}

\begin{abstract}
We present measurements of the superconducting coherence length $\xi$ in thin ($d \leq 100$ \AA) films of MoGe alloy and Nb using a combination of linear and nonlinear mutual inductance techniques.  As the alternating current in the drive coil is increased at fixed temperature, we see a crossover from linear to nonlinear coupling to the pickup coil, consistent with the unbinding of vortex-antivortex pairs as the peak pair momentum nears $\hbar/\xi$ and the unbinding barrier vanishes.  We compare measurements of $\xi$ made by this mutual inductance technique to values determined from the films' upper critical fields, thereby confirming the applicability of a recent calculation of the upper limit on a vortex-free state in our experiment.  
\end{abstract}

\maketitle

\section{Introduction}
\par The two-coil experiment is a long-established method for measuring the superfluid density $n_s \propto 1/\lambda^2$ in thin films\citep{hebfiory}, where $\lambda$ is the London penetration depth.  A common geometry is a pair of coaxial coils located on opposite sides of the film sample.  The drive coil is located closer to the film's surface, and energized with an alternating current. Meissner currents flow within the film to screen the small applied magnetic field from the drive coil.  Coupling to the second (pickup) coil is measured as a function of temperature.  Provided that the induced screening current in the film is far below its critical value, the order parameter is spatially uniform and the film can be described by a single complex conductivity.  The imaginary part of this conductivity, deduced from the linear coupling between the coils, is used to calculate the superfluid density as a function of temperature \cite{stefancalc,stefancalc2}.

\par The present work is preceded by that of Scharnhorst \cite{scharnhorst} and Claassen \cite{claassen} in which experimental critical currents were deduced from the onset of nonlinear (third harmonic) response as the drive coil current was increased for metallic and cuprate films, respectively.  Our approach expands this work in that we measure the fundamental response (amplitude and phase) over four decades of driving current, revealing heretofore un-examined features such as the unbinding of vortex-antivortex (V-aV) pairs, and hysteresis due to vortex pinning.

\par Using a simple model of our experiment, Lemberger and Draskovic\cite{LDtheory} showed that milligauss fields can sustain equilibrium V-aV pairs, assuming that there is zero free energy barrier to their unbinding.  As such, all two-coil measurements whether linear or nonlinear are made with the film in a metastable Meissner state governed by the unbinding barrier.  Lemberger and Ahmed\cite{LAtheory} calculated the upper limit of this metastable state as a function of the film's penetration depth.  They found that
\begin{equation}
B_0^{crit} \approx \Phi_0/2R\xi
\end{equation} 
for films in the limit of long 2D penetration depth $\Lambda \equiv 2\lambda^2/d > R$, where $d$ is the film thickness, and $R$ is the radial position on the film where the applied perpendicular field changes sign.  In the limit of short penetration depth $\Lambda \ll R$, they found:
\begin{equation}
B_0^{crit} \approx \Phi_0/2\pi\Lambda\xi.
\end{equation}

\par As a practical matter, our goal is to measure the coherence length in ultrathin cuprate films, where determining $\xi$ from the upper critical field $B_{c2}$ is problematic.  Our nonlinear measurements are easily performed at $T/T_c \ll 1$, and the vortex physics appears at tiny magnetic fields ($B_0<10$ G), where $B_0$ is the largest perpendicular field applied to the film, namely, the field applied at the center of the film.

\section{Experiment}

\par Thin films of amorphous molybdenum-germanium ($a$-MoGe) with thicknesses of 40 \AA, 50 \AA, 60 \AA, and 100 \AA\ (measured by calibrated deposition) were RF sputtered onto 8x8 mm$^2$ SiO$_2$-capped silicon substrates with an average rate of 0.45 \AA/s, giving $T_c$'s in a range from 3-6 K.  Nb films with $T_c$'s in a range from 2-7 K were prepared on 15x15 mm$^2$ substrates at 1.5 \AA/s.  The thicknesses of the Nb films, (19 \AA, 41 \AA, 54 \AA, and 62 \AA),  were determined by an empirical fit of $T_c(d)$ determined previously \cite{tomnb} for films of identical preparation. The Nb films received a cap layer of several hundred \AA's of Ge sputtered at 2 \AA/s to prevent oxidation of the film between growth and characterization. 

\par
\begin{figure}
\includegraphics[scale=0.60]{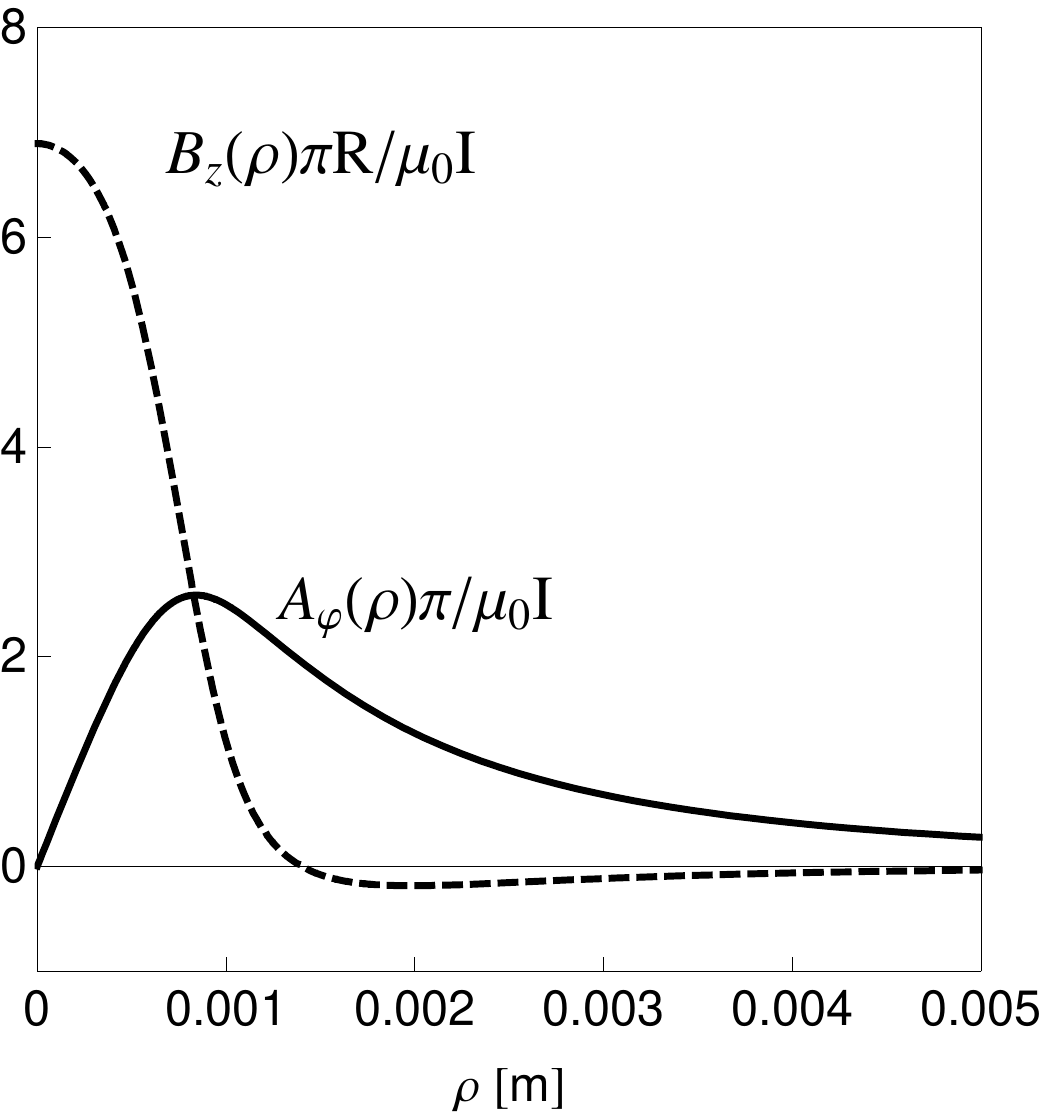}
\caption{Perpendicular magnetic field (dashed) and the vector potential (solid) produced by the experimental drive coil.  The coil, radius 660 $\mu$m, consists of 16 turns of Nb-Ti wire around a plastic form and located 380 $\mu$m from the film surface.}
\label{fig:coilfield}
\end{figure}

 Our coils are wound from Nb-Ti wire with $T_c \approx 9$ K, higher than the highest-$T_c$ films in this study. The applied perpendicular field, $B_z(\rho)$ and in-plane vector potential, $A_\phi(\rho)$, from the drive coil are displayed in Figure \ref{fig:coilfield}. $R \approx$ 1.4 mm is the radial distance at which $B_z$ changes sign, a little more than double the coil radius. A commercial audio amplifier provided 10 kHz sinusoidal drive currents ranging from $<$50 $\mu$A to 1 A.  Drive coil current and voltage across the pickup coil were measured by lock-in amplifiers.

\par Measurements were made according to the following schedule for each film:  The sample was cooled to 1.4 K in our two-coil cryostat.  Nonlinear measurements were conducted at 1.4 K.  Linear measurements were conducted as the two-coil cryostat warmed.  The sample was then transferred to a PPMS\textsuperscript{\textregistered} (the Nb films were cut down to fit the sample holder) and cooled again to measure the resistance transition.  High magnetic field sweeps were made at fixed low temperatures, and the sample was warmed again. Table \ref{table:data} summarizes the measurements made on the eight films.

\begin{table}
\setlength{\tabcolsep}{5pt}
\caption{Measurements made on four MoGe films and four Nb films.  $T_{c,\rho}$ is the transition temperature as measured by four-terminal resistance.  $T_{c,0}$ and $\Lambda(0) \equiv 2\lambda^2/d \propto 1/n_s(0)$ are taken from a dirty-limit BCS fit to the superfluid density data.  $\xi_{Bc2}(0) \equiv \sqrt{\Phi_0/2 \pi B_{c2}(0)}$.}
\label{table:data}
\begin{center}
\begin{tabular}{c c c c c c c}
\hline
\hline 
{} & $d$& $T_{c,\rho}$& $T_{c,0}$& $\Lambda(0)$ & $B_{NL}(1.4 K)$ & $\xi_{Bc2}(0)$ \\
{} & (\AA) & (K) & (K) & ($\mu$m) & (G) & (\AA) \\  
\hline
MoGe & 40 & 3.6 & 3.55 & 312 & 2.8 & 71 \\
{} & 60 & 4.9 & 5.00 & 133 & 7.2 & 69 \\
{} & 80 & 5.6 & 5.55 & 71 & 7.5 & 55 \\
{} & 100 & 5.3 & 5.46 & 68 & 8.3 & 56 \\
\hline
Nb & 19 & 2.6 & 2.22 & 505 & 0.72 & 92 \\
{} & 41 & 4.5 & 4.45 & 65 & 2.4 & 105 \\
{} & 54 & 5.8 & 5.50 & 16 & 8.6 & 107 \\
{} & 62 & 6.4 & 6.00 & 10 & 14.4 & 110 \\
\hline
\hline 
\end{tabular}
\end{center}
\end{table}

\subsection{Linear measurements}
\par
\begin{figure}
\includegraphics[scale=0.75]{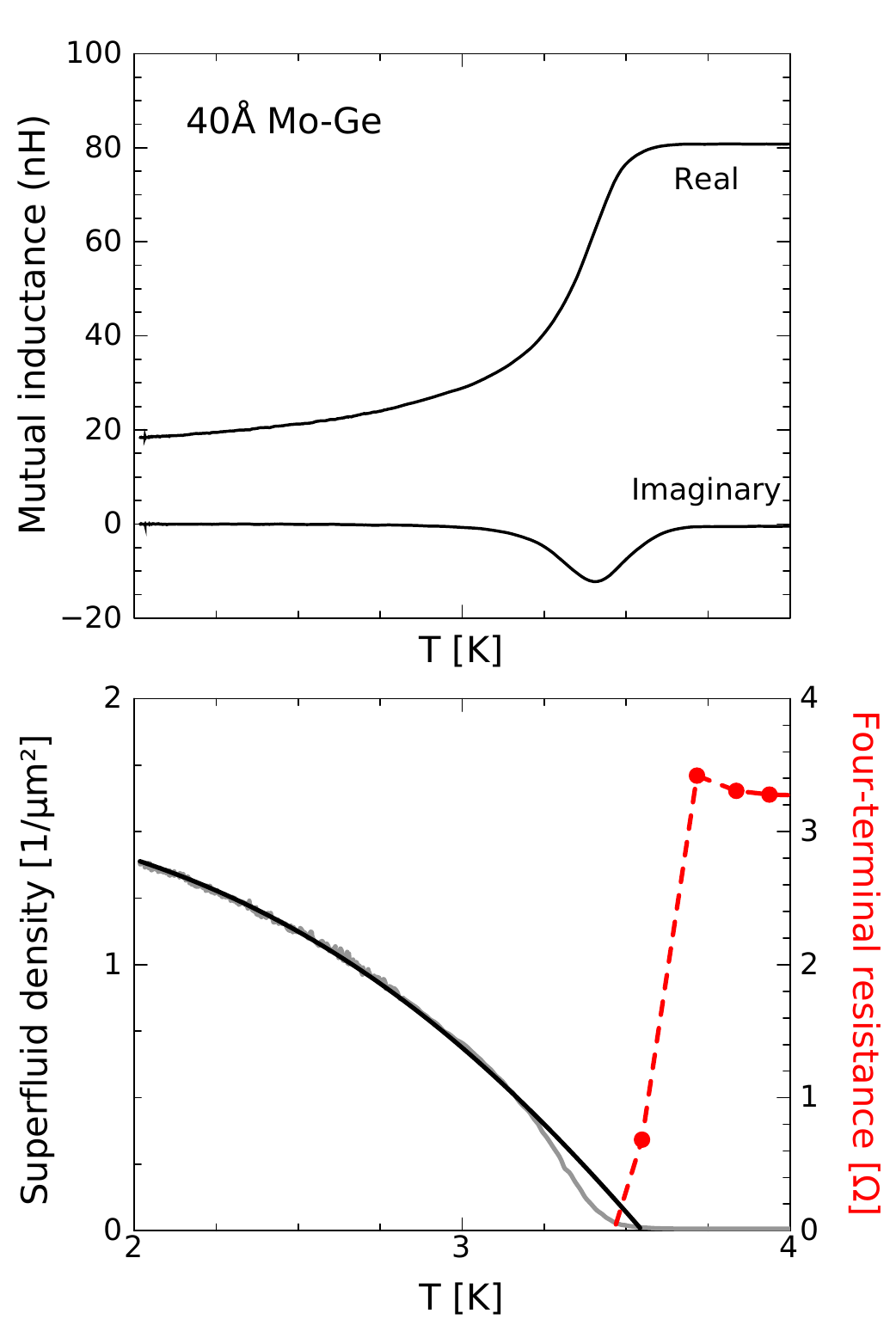} \\ 
\begin{center} 
\caption{Complex mutual inductance as a function of temperature in the linear regime for a typical MoGe film (\textit{top}).  Upper curve is the real part of the mutual inductance $M = V_{p}/ \omega I_{d}$, where $V_p$ is the out-of-phase component of the pickup coil voltage and $I_d$ is the drive coil current.  Lower curve is the imaginary part. \textit{Bottom:} Superfluid density and four-terminal resistance as a function of temperature for the same film.  Black line is a dirty-limit BCS fit to data (\textit{in grey}).}
\label{fig:superfluid}
\end{center}
\end{figure}

\begin{figure}
\includegraphics[scale=0.75]{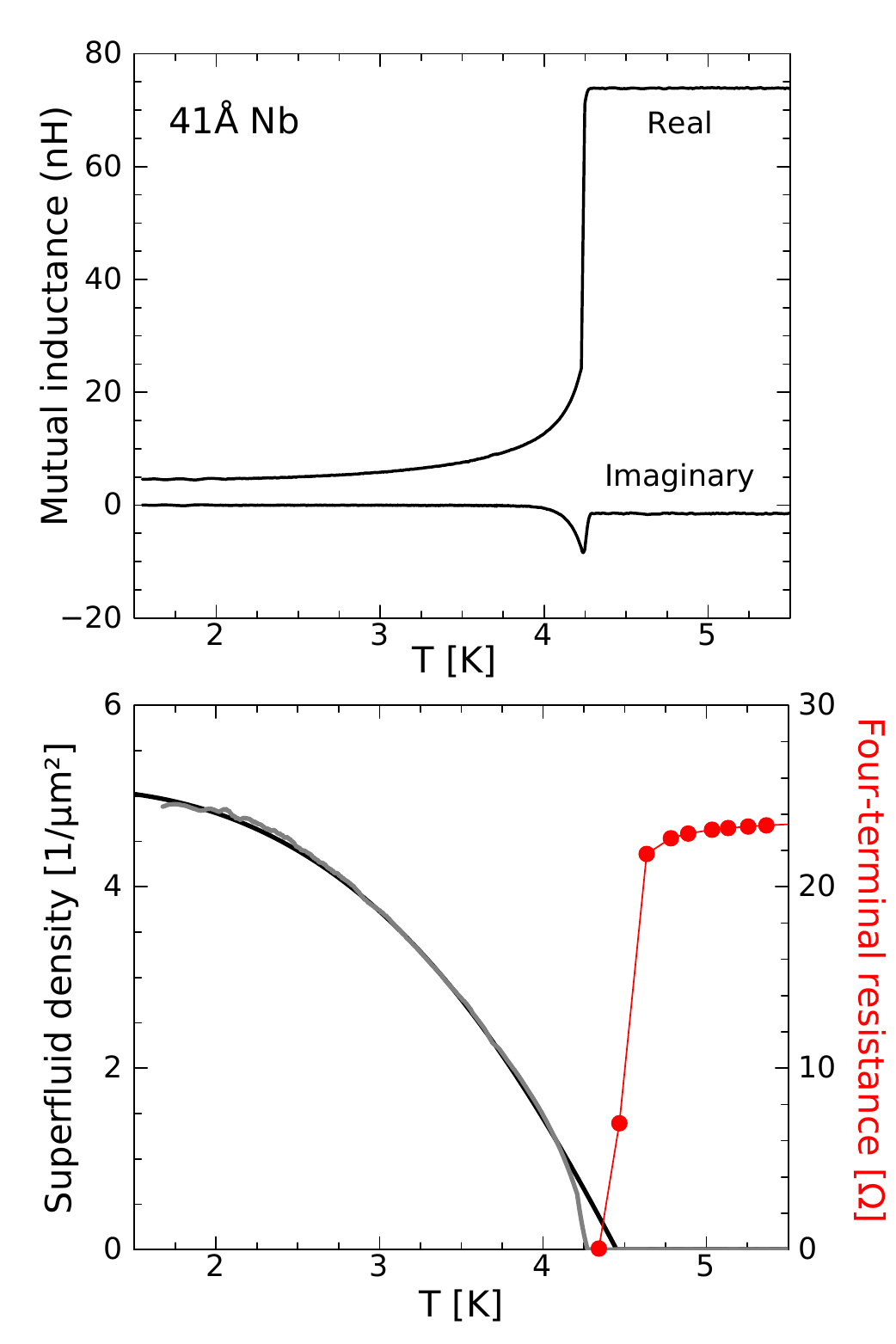}\\
\begin{center} 
\caption{Complex mutual inductance as a function of temperature in the linear regime for a typical Nb film (\textit{top}).  \textit{Bottom:} Superfluid density and four-terminal resistance as a function of temperature for the same film.}
\label{fig:superfluid2}
\end{center}
\end{figure}

 To determine the magnetic penetration depths, the alternating current in the drive coil was held well below the non-linear crossover (i.e., $B_0 < 5$ mG), and the cryostat was allowed to warm up.  Superfluid density data were fit to the BCS dirty-limit per Ref.'s \onlinecite{tomnb} and \onlinecite{stefanmoge} with a gap parameter $\Delta(0)/k_B T_c = 1.9$. Typical data for mutual inductance, corresponding superfluid densities $\lambda^{-2}(T)$, and 4-terminal resistances for $a$-MoGe and Nb films are shown in Figures \ref{fig:superfluid} and \ref{fig:superfluid2}. Note that the superfluid density diminishes below the mean-field $T_{c,0}$ for these films. This is also evident in the appearance of nonzero resistance below the mean-field transition. We attribute this to currents in the film exceeding the critical current near the transition. We believe that this deviation in superfluid would diminish if suitably small drive/excitation currents had been used.  We are confident that our experimental values of $1/\Lambda(T \rightarrow 0)$ are accurate to $\pm$10\%. $1/\Lambda$ is measured directly, so uncertainty in film thickness is irrelevant.\cite{stefancalc}$^,$\cite{stefancalc2}

\subsection{Nonlinear Measurements}
\par
\begin{figure}
\begin{tabular}{c}
\includegraphics[scale=0.40]{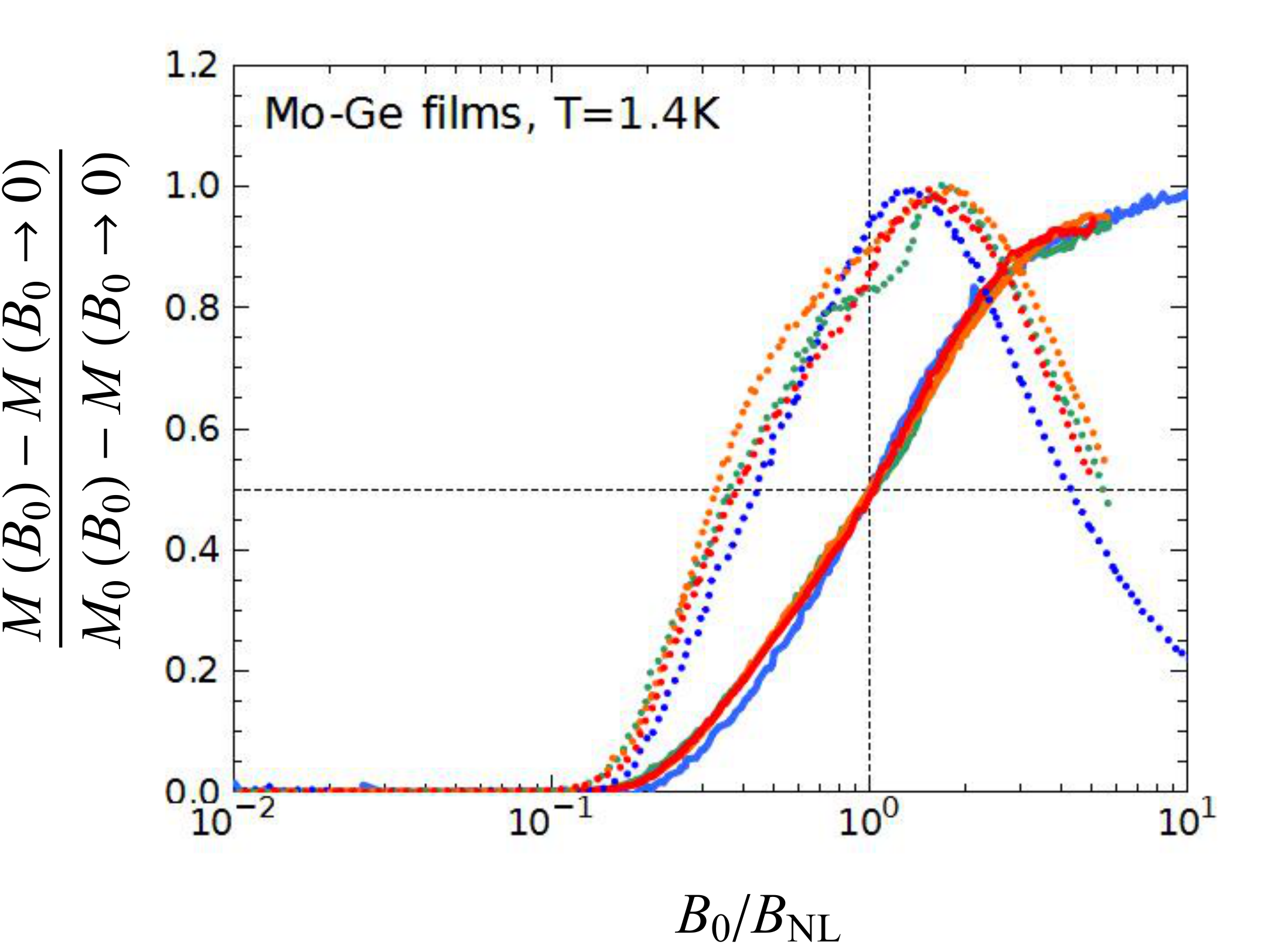}\\
\includegraphics[scale=0.40]{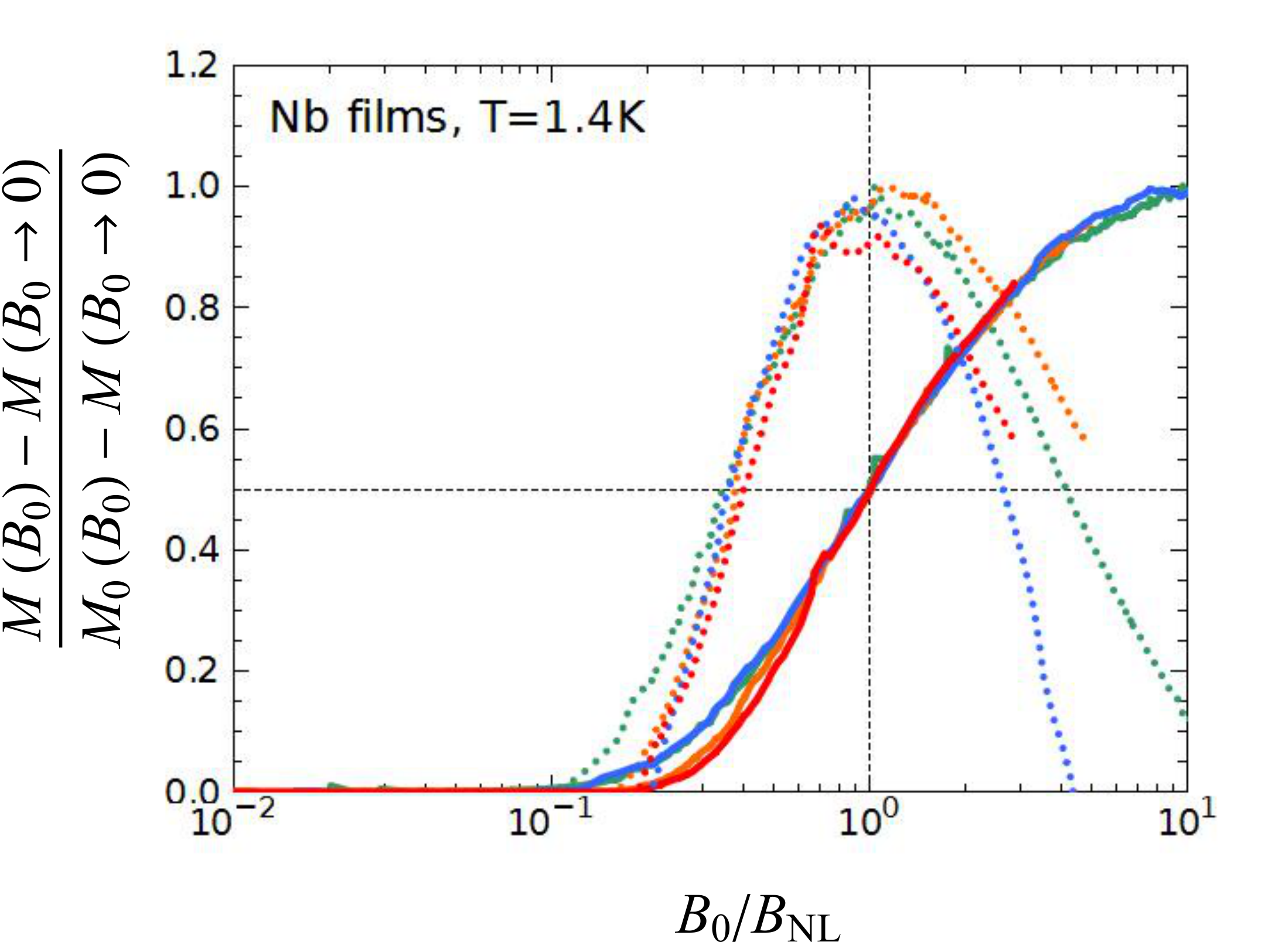}\\
\end{tabular}
\caption{
%Complex mutual inductance as a function of drive coil current for the 40\AA -thick MoGe film (\textit{top left}).\textit{Top:} 
Scaled complex mutual inductance of four MoGe films and four Nb films ($M_0$ is the mutual inductance when the film is in the normal state) as a function of the applied magnetic field amplitude at the center of the film $B_0$, scaled by the applied field where the mutual inductance has reached half of its normal-state value, $B_{NL}$ (\textit{solid curves}).  Dotted curves are the absolute value of imaginary mutual inductance normalized to its peak value.}
\label{fig:nldata}
\end{figure}

 Nonlinear measurements were conducted with the experimental probe submerged in a LHe bath at the lowest temperature achievable by pumping on the bath, approximately 1.4 K.  We feel that the combination of superconducting coils and immersion in LHe is sufficient to nullify the coil heating concerns raised by Refs. \onlinecite{scharnhorst} and \onlinecite{claassen}, where non-superconducting coils were used. Samples were cooled to 1.4 K in liquid helium with the drive coil de-energized. The drive coil current was increased to produce the mutual inductance curves
of Fig. \ref{fig:nldata}. 

\par The linear response regime, clearly indicated by a field-independent value of the real part of mutual inductance, crosses over to a regime where the mutual inductance rises monotonically toward its normal-state value, $M_0$. We characterize this effect by the value of $B_0$ at which the mutual inductance has risen halfway to its normal state value, denoting this field as $B_{NL}$. The similarity of the scaled curves demonstrates the consistency of this effect across films of different thickness and composition.  We note that these curves are only very weakly dependent on frequency in the range of 1-20 kHz where our electronics are linear, and postpone a thorough exploration of the frequency domain.

\subsection{High-field measurements}
\par
\begin{figure}
\includegraphics[scale=0.65]{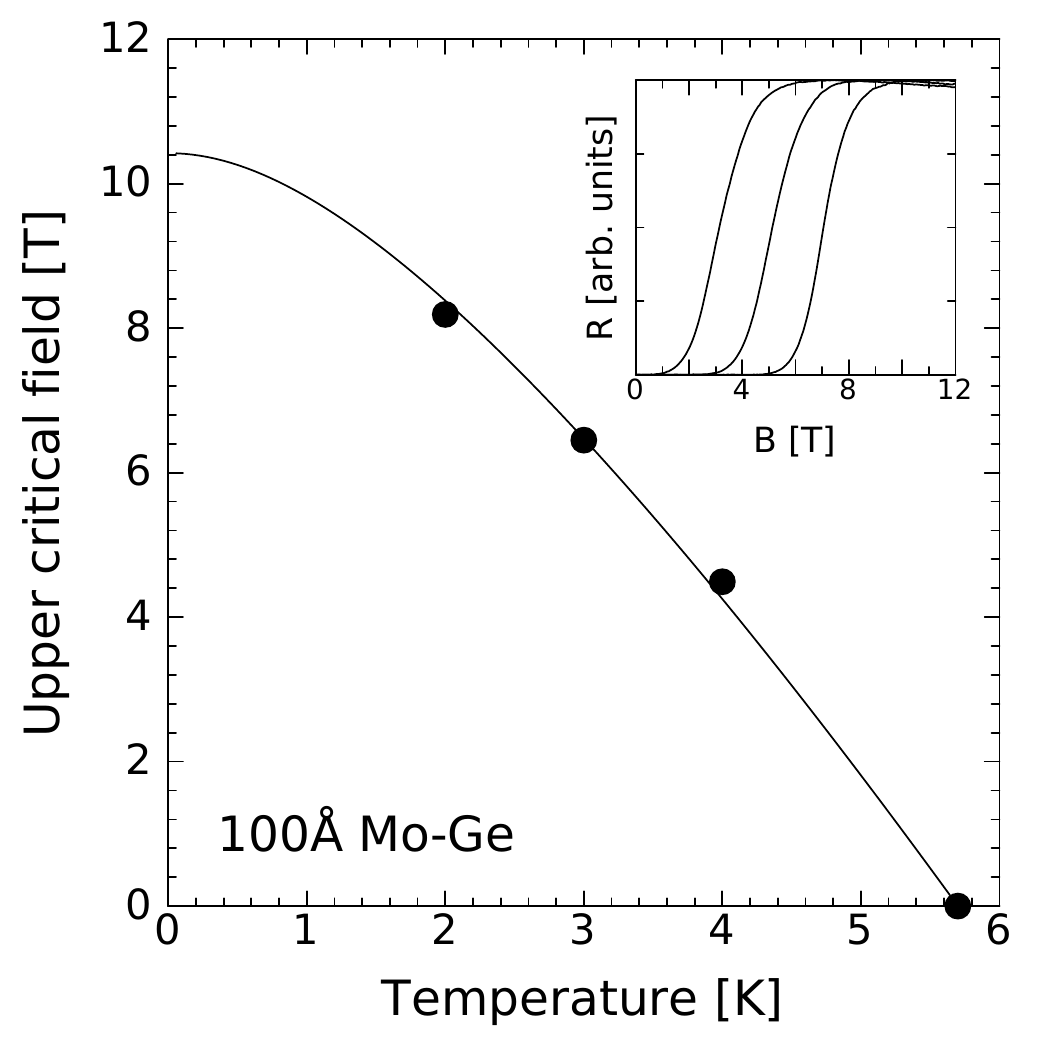}
\caption{Upper critical field $B_{c2}$ as a function of temperature for a MoGe film.  $B_{c2}(T)$ is determined by the measured resistance reaching 90\% of its normal-state value for constant temperature data, as well as constant field data for $B=0$.  Fit is Abrikosov-Gorkov curve.}
\label{fig:bc2data}
\end{figure}

\begin{figure}
\includegraphics[scale=0.65]{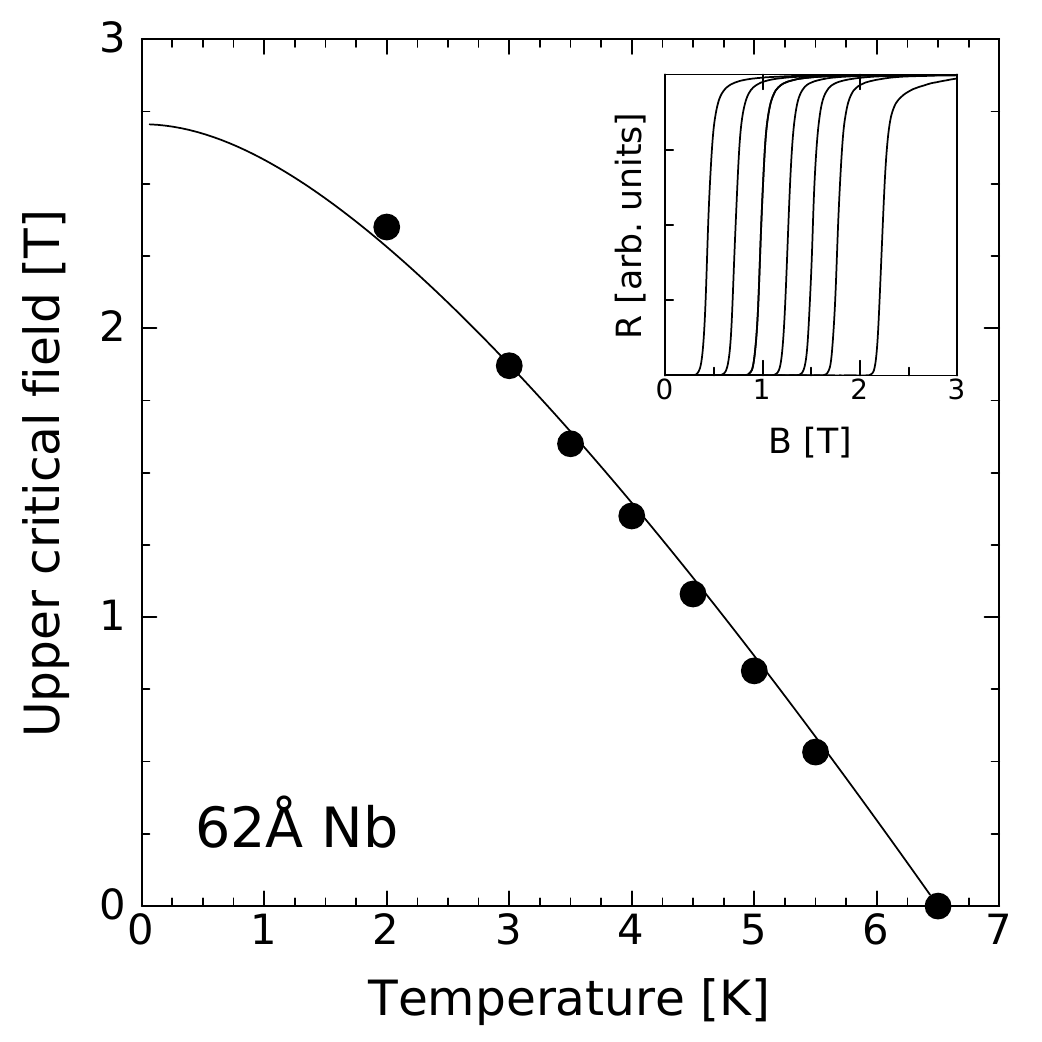}
\caption{Upper critical field $B_{c2}$ as a function of temperature for a Nb film.}
\label{fig:bc2data2}
\end{figure}

 We measured perpendicular upper critical fields $B_{c2}(T)$ from resistive transitions, \textit{$R_{sheet}$ vs. B}, in fields up to 14 Tesla in a Quantum Designs PPMS\textsuperscript{\textregistered}, as shown in Figures \ref{fig:bc2data} and \ref{fig:bc2data2}. Samples were wired in a Van der Pauw configuration. To determine $\xi(0)$, $B_{c2}(T)$ was extrapolated to zero temperature by fitting to the dirty-limit Abrikosov-Gor'kov form.\cite{hftheory} We found that determining $B_{c2}(T)$ by a four-terminal resistance value of 90\% of the normal-state value gave $B_{c2}(0)$ values consistent with textbook data\citep{bezA}$^,$\citep{nbchap}.  Uncertainty in $\xi_{Bc2}$ is set by the width of the resistive transitions (\textit{see inset to Figures \ref{fig:bc2data} and \ref{fig:bc2data2}}). The broadest transitions produce an uncertainty of $\pm 25$\% in $B_{c2}$, but given the inverse-square relationship between $B_{c2}$ and $\xi$, the error in $\xi$ is halved.

\par Dirty limit BCS theory is applicable if the mean free path $l$ is less than the coherence length.  In amorphous MoGe, $l < 10$ \AA\ (a few interatomic spacings) for thick films\cite{graybeas}.  Since our films are thicker than 10\AA, surface scattering should not be important.  The dirty limit gives $\xi \propto \sqrt{\hbar v_f l/\pi \Delta(0)}$, so for constant $l$ and Fermi velocity we expect
\begin{equation}
\xi = \xi^{T_c=7.3K}\sqrt{\frac{7.3 K}{T_c}}, \hspace{1 cm} \text{(MoGe films)}
\end{equation}
which fits $\xi_{Bc2}$ well with $\xi_{Bc2}^{T_c=7.3K}$ = 65 \AA\ for MoGe.

\par In thin Nb, the mean free path \textit{is} limited by surface scattering\cite{tomnb}: $l \approx d/4$.  Since $T_c \propto d$ for low-$T_c$ films, we expect
\begin{equation}
\xi = \xi^{T_c=8.5 K} \approx \text{Const.} \hspace{1 cm} \text{(Nb films)}
\end{equation}
The mean of our $B_{c2}$-derived values for Nb is 103 \AA\ $\pm$ 8\% --- constant with respect to our experimental uncertainty.

%\par Since $\Lambda \gg d$ for our films, Meissner screening currents are uniform through the thickness of a film in both the linear and nonlinear regimes. Thus, our films are effectively two-dimensional. Lateral variation in screening current density is set by $R$ = 1.4 mm.    Lateral variations in supercurrent density associated with 2D vortices\cite{pearl} are at the scale of $\Lambda < R/2$ for each of our films.

\section{Discussion}

\par The similarity of the curves in Figure \ref{fig:nldata} from film-to-film and between Nb and $a$-MoGe argues that the basic physics is largely the same in all films, regardless of differences in disorder, vortex pinning details, etc.  A simple critical state model of the vortex behavior based on a "puddle" of vortices forming at the center of the film under each cycle of the driving field captures the qualitative features of the experimental data, namely a rise in the real mutual inductance coincident with a peak in the imaginary mutual inductance signal.  The imaginary mutual inductance is a result of hysteresis in the areal densities of vortices and antivortices, a feature that is not seen unless pinning is included in the model.  The model calculations, details of which are the subject of a forthcoming manuscript, indicate that the real mutual inductance rises with near-vertical slope and reaches 50\% of the normal state value within a factor of two of the field of first vortex-antivortex unbinding at zero temperature.  The peak in the calculated imaginary mutual inductance occurs simultaneously.  Given the plausibility of V-aV pairs unbinding thermally in our finite-temperature experiment, we take $B_{NL}$ to correspond to the bulk unbinding of vortex-antivortex pairs in the data, rather than the field at which the signal first deviates from linearity, approximately $B_{NL}/5$.

\par
\begin{figure}
\includegraphics[scale=0.35]{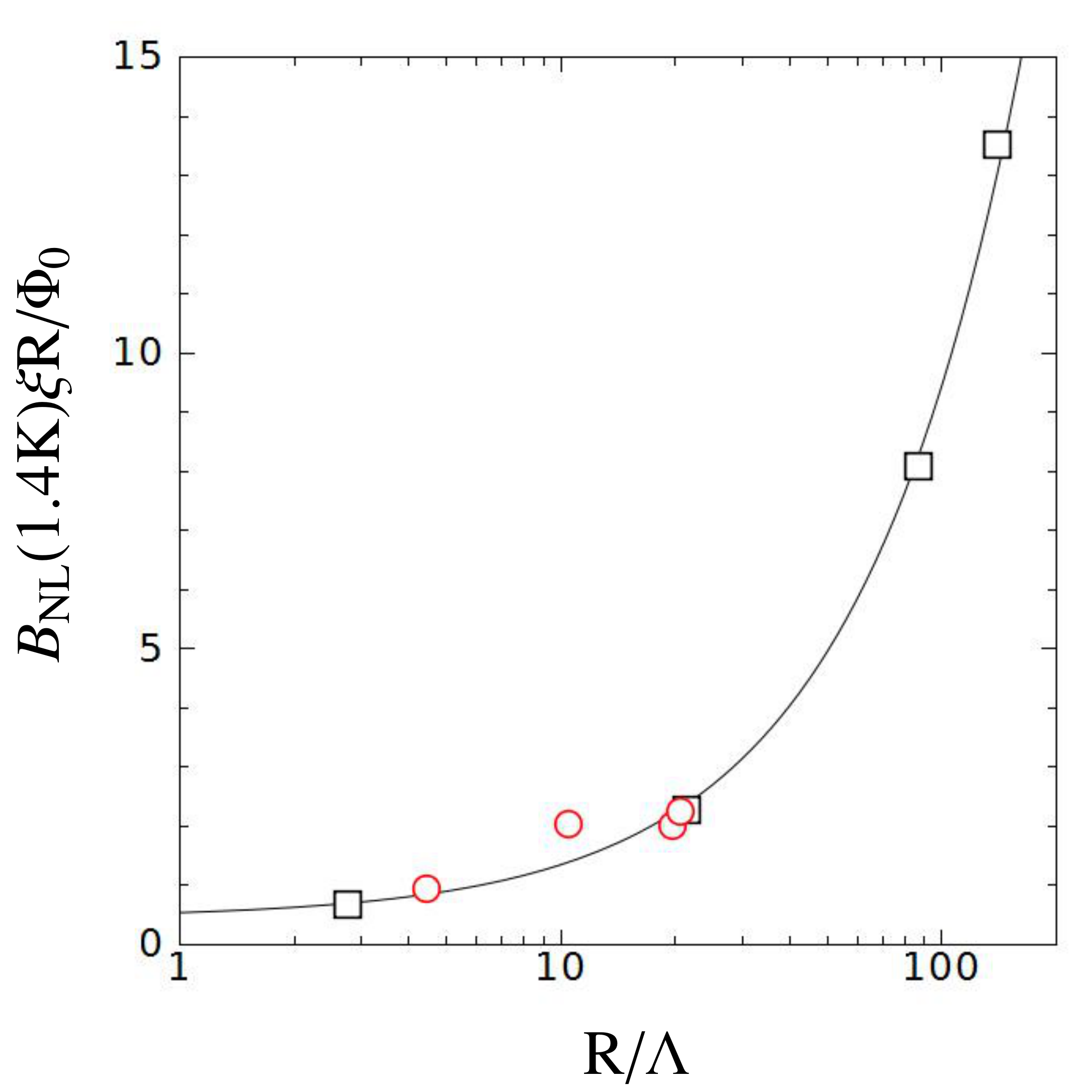}
\caption{Normalized $B_{NL}$ measurements plotted as a function of dimensionless superfluid density $R/\Lambda$, and fitted to the single-ring model of Lemberger and Ahmed\cite{LAtheory}: $y \approx 0.45+0.09x$.  $\xi$ for each material is the only fitting parameter, as per Eq. 3 and 4.  The fit gives $\xi^{T_c = 7.3 K}$ = 33 \AA\ for MoGe films (red circles) and $\xi^{T_c = 8.5 K}$ = 135 \AA\ for Nb films (black squares).}
\label{fig:allfilms}
\end{figure}

 In Figure \ref{fig:allfilms}, the $B_{NL}$ values of Table \ref{table:data} are fitted to the model of Ref. \onlinecite{LAtheory} (Eq. 1 and 2) using the functional forms of Equations 3 and 4 with $\xi^{T_c = 7.3 K}$ and $\xi^{T_c = 8.5 K}$ as free parameters for MoGe and Nb, respectively. This amounts to a qualitative test of the model, and the fit is excellent.  The model gives quantitative agreement as well: the fit parameters used in Figure \ref{fig:allfilms} give factor-of-two agreement with the $B_{c2}$-determined values.

\section{Summary}
We have measured the complete crossover of linear to nonlinear two-coil response for two well-understood dirty-limit BCS superconductors.  We have empirically identified the applied driving magnetic field associated with vortex-antivortex pairs unbinding \textit{en masse} in the films, informed by a microscopic model which includes flux pinning.  Our data on eight films fit well to Lemberger and Ahmed's recent calculation of the upper bound of the vortex-free state as a function of superfluid density\cite{LAtheory}, and we are able to extract values of the superconducting coherence length from these data in quantitative agreement with values measured from the films' upper critical fields.

\begin{acknowledgments}
This work was supported in part by DOE-Basic Energy Sciences through Grants No. FG02-08ER46533 and DE-FG0207ER46453, and in part by NSF Grants DMR-0805227 and DMR 10-05645.

\end{acknowledgments}

\bibliography{references}

\end{document}